\pdfoutput=1
\documentclass[sigconf, nonacm]{acmart}
\makeatletter
\let\@acmBadgeL@image\@empty
\let\@acmBadgeR@image\@empty
\makeatother

\makeatletter
\@ifundefined{correspondingauthor}{\newcommand\correspondingauthor{}}{}
\makeatother

\usepackage{booktabs}
\usepackage{multirow}
\usepackage{graphicx}
\usepackage{tikz}
\usetikzlibrary{positioning, arrows.meta}
\newcommand\vldbyear{2026}
\newcommand\vldbworkshop{QC\&DKM 2026- 2nd Workshop on Quantum Computing and Data/Knowledge Management}
\newcommand\vldbauthors{\authors}
\newcommand\vldbtitle{\shorttitle} 
\newcommand\vldbavailabilityurl{https://github.com/umich-db/quantumdb/tree/spiq-joo}
\newcommand\vldbpagestyle{plain}

\definecolor{lin}{HTML}{e31a1c}

\begin{document}
\title{Improving Join Order Optimization on Gate-Based Quantum Computers via Structured Parameter Initialization}


\author{Divya Shekar}
\authornote{Both authors contributed equally to this research.}
\email{shdivya@umich.edu}
\author{Ruokun Wu}
\correspondingauthor
\authornotemark[1]
\email{ruokun@umich.edu}
\affiliation{%
  \institution{University of Michigan}
  \city{Ann Arbor}
  \state{Michigan}
  \country{United States}
}



\author{Dhanvi Bharadwaj}
\affiliation{%
  \institution{University of Michigan}
  \city{Ann Arbor}
  \state{Michigan}
  \country{United States}
}
\email{dhanvib@umich.edu}

\author{Gokul Subramanian Ravi}
\affiliation{%
  \institution{University of Michigan}
  \city{Ann Arbor}
  \state{Michigan}
  \country{United States}
}
\email{gsravi@umich.edu}

\author{Lin Ma}
\affiliation{%
  \institution{University of Michigan}
  \city{Ann Arbor}
  \state{Michigan}
  \country{United States}
}
\email{linmacse@umich.edu}


\begin{abstract}
Join Order Optimization (JOO) is one of the most computationally expensive tasks in relational query optimization due to the exponential growth of possible join plans with increasing query size. Recent work has explored quantum and quantum-inspired approaches for solving JOO by reformulating the problem as a Quadratic Unconstrained Binary Optimization (QUBO) problem suitable for optimization using quantum hardware. However, many existing approaches have limited scalability on current gate-based quantum devices. In addition, little work has investigated the role of initialization strategies in improving the performance of gate-based quantum optimization for database workloads. In this work, we investigate gate-based quantum join order optimization using the Quantum Approximate Optimization Algorithm (QAOA) initialized with Scalable Parameter Initialization for QAOA (SPIQ). SPIQ  is used to efficiently identify high-quality initial points in the quantum solution landscape for QAOA executed on a gate-based quantum computer. We evaluate the interaction between QUBO encoding, SPIQ initialization, and gate-based optimization on small-scale join ordering problems involving 3 and 4 relations. Our results show that structured initialization improves optimization stability and increases convergence toward high-quality join plans compared to uninformed initialization approaches. Across these small-scale, simulation-based instances, SPIQ increases the sampling frequency of the optimal join order by up to approximately 5$\times$ and yields final-state energies significantly lower than a randomly initialized QAOA. Overall, this work enhances existing gate-based quantum optimization while providing an initial proof of concept for applying SPIQ initialization to database query optimization workloads.

\end{abstract}

\maketitle
\pagestyle{\vldbpagestyle}
\begingroup\small\noindent\raggedright\textbf{VLDB Workshop Reference Format:}\\
\vldbauthors. \vldbtitle. VLDB \vldbyear\ Workshop: \vldbworkshop.\\ 
\endgroup
\begingroup
\renewcommand\thefootnote{}\footnote{\noindent
This work is licensed under the Creative Commons BY-NC-ND 4.0 International License. Visit \url{https://creativecommons.org/licenses/by-nc-nd/4.0/} to view a copy of this license. For any use beyond those covered by this license, obtain permission by emailing \href{mailto:info@vldb.org}{info@vldb.org}. Copyright is held by the owner/author(s). Publication rights licensed to the VLDB Endowment. \\
\raggedright Proceedings of the VLDB Endowment. 
ISSN 2150-8097. \\
}\addtocounter{footnote}{-1}\endgroup

\ifdefempty{\vldbavailabilityurl}{}{
\vspace{.3cm}
\begingroup\small\noindent\raggedright\textbf{VLDB Workshop Artifact Availability:}\\
Github Codebase: \url{https://github.com/umich-db/quantumdb/tree/spiq-joo}. 
\endgroup
}

\section{Introduction}
The performance of modern Relational Database Management Systems (RDBMS) is fundamentally dependent on the quality of query execution plans generated by the optimizer~\cite{selinger1979access}. Among the many components of query planning, Join Order Optimization (JOO) remains one of the most computationally challenging and performance-critical tasks~\cite{steinbrunn1997heuristic,moerkotte2006analysis}. Given a query involving multiple relations, the optimizer must determine an execution order that minimizes intermediate result sizes and overall execution cost. The number of possible join orders grows exponentially with the number of relations, making the problem NP-hard and increasingly difficult for large enterprise workloads involving tens or hundreds of tables~\cite{ioannidis1996query}. Classical optimizers traditionally rely on dynamic programming, greedy heuristics, and cost-based pruning strategies using relation statistics and predicate selectivities~\cite{selinger1979access,graefe1993query}. However, as modern workloads continue to grow in scale and complexity, classical hardware and exhaustive search methods struggle to efficiently explore the rapidly expanding solution space.

Quantum computing has emerged as a promising alternative paradigm for solving large combinatorial optimization problems~\cite{nielsen2010quantum,preskill2018quantum}. Unlike classical computation, which evaluates candidate solutions sequentially, quantum systems exploit superposition and entanglement to explore multiple states simultaneously. This has motivated the application of quantum algorithms to database optimization problems, particularly join ordering. By representing join order optimization as an energy minimization problem, candidate execution plans can be encoded into a Hamiltonian whose ground state corresponds to the minimum-cost join order. Recent work has demonstrated encouraging proof-of-concept results using Quantum Approximate Optimization Algorithm (QAOA)~\cite{farhi2014quantum}, quantum annealing~\cite{kadowaki1998quantum}, and quantum-inspired optimization frameworks~\cite{lucas2014ising}. These approaches suggest that quantum computing may eventually provide scalable mechanisms for navigating the exponential search space of query optimization more efficiently than classical techniques.

Most prior work in quantum join order optimization has focused on annealing-based or quantum-inspired optimization methods rather than gate-based quantum computation~\cite{schonberger2023quantum}. While such approaches have demonstrated favorable scalability and reduced encoding complexity, they do not capture the full expressive power of universal quantum computation. Gate-based quantum computing, by contrast, provides a fundamentally quantum framework that supports parameterized quantum circuits, quantum interference, and variational algorithms such as QAOA~\cite{farhi2014quantum,cerezo2021variational}. Gate-based methods are therefore both more general and theoretically more powerful than annealing-based systems, making them an important direction for long-term quantum database research.

Despite their theoretical advantages, gate-based quantum algorithms remain difficult to optimize in practice. Current Noisy Intermediate-Scale Quantum (NISQ) devices are constrained by limited qubit counts, short coherence times, restricted qubit connectivity, and high noise levels~\cite{preskill2018quantum}. In addition, QAOA itself introduces a challenging hybrid optimization problem in which the quality of the final solution depends heavily on the initialization and optimization of circuit parameters~\cite{zhou2020quantum,cerezo2021variational}. Poor initialization can cause the optimizer to converge to low-quality local minima or produce unstable training behavior, especially for highly non-convex energy landscapes such as those arising in join order optimization~\cite{mcclean2018barren}. 

To address this issue, recent work has explored improved initialization strategies for variational quantum algorithms. Techniques such as layer-wise initialization~\cite{zhou2020quantum}, transfer-based initialization, warm-start methods~\cite{egger2021warm}, and SPIQ (Scalable Parameter Initialization for QAOA)~\cite{bharadwaj2026spiq} aim to provide parameter values closer to promising regions of the optimization landscape before training begins. By improving the starting point of the variational optimization process, these methods can potentially accelerate convergence and increase the probability of reaching near-optimal solutions on noisy hardware.

In this work, we investigate the application of SPIQ initialization to gate-based quantum join order optimization using QAOA, represented by a QUBO. By using the compact native encoding from ~\cite{schonberger2023quantum}, we require only a reasonable number of program qubits. The resulting QUBO is then mapped to an Ising Hamiltonian and optimized using QAOA initialized with SPIQ-generated parameters. We evaluate using noiseless statevector simulation to isolate the impact of SPIQ initialization on optimization without hardware noise effects. While noise-aware simulations could provide insight into real-device performance, they primarily reflect hardware fidelity rather than initialization quality and incur significantly higher computational cost. Evaluating SPIQ under realistic noise models and on physical hardware remains future work.

Our evaluation focuses on small-scale join queries over 3 and 4 relations, with problem size constrained by the cost of classical simulation of quantum systems. Within this regime, we analyze the interaction between the QUBO encoding, SPIQ initialization, and gate-based QAOA optimization. The results show that informed initialization improves convergence toward high-quality candidate join orders and yields more stable optimization behavior than uninformed initialization. With SPIQ-initialized parameters, we achieve at most a 5$\times$ improvement in optimal cost frequency, and the final state energy is significantly lower than that of the randomly initialized QAOA across the three problems. We expect these benefits to persist at larger problem sizes when executed on real quantum hardware, though qubit error effects will need to be accounted for. 
 The primary contributions of this work are as follows: (1) adapting the encoding from annealing-oriented quantum JOO approaches to gate-based quantum optimization for better scalability, (2) presenting the first application of SPIQ initialization to gate-based quantum join order optimization, and (3) identifying practical limitations and future research directions for scalable quantum-enhanced query optimization.

\section{Background and Existing Work}

\subsection{Quantum Computing}

Quantum computing is an emerging computational paradigm that utilizes quantum mechanical phenomena such as superposition, entanglement, and interference to solve computational problems~\cite{nielsen2010quantum,preskill2018quantum}. Unlike classical bits, which exist only in states 0 or 1, quantum bits (qubits) can exist in linear combinations of both states simultaneously. This enables quantum systems to explore large solution spaces more efficiently than classical architectures for certain classes of optimization and search problems. Current quantum computing approaches can broadly be divided into two categories: gate-based quantum computing and quantum annealing.

Gate-based quantum computing operates using sequences of parameterized quantum gates applied to qubits in a circuit model. Universal quantum computers belong to this category and support algorithms such as Grover's search algorithm~\cite{grover1996fast}, Variational Quantum Eigensolvers (VQE)~\cite{peruzzo2014variational}, and the Quantum Approximate Optimization Algorithm (QAOA)~\cite{farhi2014quantum}. QAOA is particularly relevant for combinatorial optimization because it solves optimization problems by encoding them into a Hamiltonian and variationally minimizing the corresponding energy function through alternating cost and mixing operators. However, current gate-based systems remain limited by noise, decoherence, restricted qubit connectivity, and small qubit counts, making large-scale optimization difficult on present-day Noisy Intermediate-Scale Quantum (NISQ) hardware~\cite{preskill2018quantum,cerezo2021variational}.


\subsection{Quantum Computing for Join Order Optimization}


The main idea in the quantum line of work for Join Order Optimization (JOO) is to rewrite join ordering as a binary optimization problem that can be solved by quantum or quantum-inspired hardware. Most existing approaches formulate the problem as either a QUBO or Ising model because these representations are compatible with both annealing-based methods and gate-based approaches such as QAOA after suitable transformation \cite{lucas2014ising,farhi2014quantum}. In practice, the primary challenge is not only to represent valid join orders, but also to encode predicate applicability and intermediate-result cost without introducing excessive binary variables or dense pairwise couplings. This tradeoff between model expressiveness and hardware feasibility is a recurring theme throughout the literature.

One of the earlier and most relevant works ~\cite{schonberger2023ready} in this area presents what the authors describe as the first implementation of join order optimization on gate-based quantum hardware. The paper emphasizes co-design between the MILP-derived encoding and hardware limitations while still preserving the semantics of join ordering. Although the work successfully demonstrated the feasibility of gate-based quantum join optimization, it also highlighted major scalability challenges, particularly the rapid increase in qubit requirements due to encoding overhead.

In a closely related follow-up \cite{schonberger2023quantum}, instead of targeting gate-based hardware, the focus is on Fujitsu's Digital Annealer, a quantum-inspired optimization platform designed for large-scale QUBO problems. The authors propose a native encoding specifically tailored for annealing-style hardware rather than adapting earlier MILP-derived formulations. The reported experiments demonstrate scalability to approximately 50 joined relations, substantially larger than the small-scale instances feasible on gate-based hardware. While this work improves practicality and scalability, it remains quantum-inspired rather than fully gate-based quantum optimization.


More recent work has also explored hybrid and system-level approaches including a hybrid quantum-classical query optimizer \cite{liu2025demonstration} integrated into a real database environment and describes itself as the first practical quantum-augmented optimizer within a working database system, constrained quadratic models, hybrid classical-quantum optimization, and alternative encoding strategies, collectively demonstrating that the field is gradually moving toward more scalable and practically usable quantum query optimization frameworks \cite{venturelli2019reverse,mohammadi2022qubo}. Another line of work is end-to-end quantum-enhanced query optimization. DEDALUS \cite{limnaios2026dedalus} presents a framework that integrates QUBO-based join order optimization with search-space pruning and cost-aware optimization, while supporting both classical and quantum optimization. 

\subsection{Initialization Strategies for Quantum Computing}

Variational quantum algorithms such as QAOA rely heavily on the initialization of circuit parameters \cite{cerezo2021variational}. Since QAOA optimizes a highly non-convex energy landscape using a hybrid classical-quantum optimization loop, poor initialization can lead to convergence toward low-quality local minima, unstable optimization behavior, or barren plateau effects where gradients vanish during training \cite{mcclean2018barren}. As a result, initialization strategies have become an important research direction for improving the practical performance of variational quantum algorithms.

In QAOA, the optimization process is controlled by sets of variational parameters, commonly denoted as $\gamma$ and $\beta$, which determine how strongly the quantum circuit applies the problem Hamiltonian and mixing operations at each layer. Traditionally, these parameters are initialized randomly, after which a classical optimizer iteratively updates them to minimize the measured energy of the quantum state. However, because the optimization landscape is highly non-linear and often contains many local minima, random initialization can lead to poor convergence behavior or require a large number of optimization iterations before useful solutions are found. Consequently, recent work has explored structured initialization techniques that attempt to place the initial parameters closer to regions of the landscape associated with lower-energy solutions.

One important class of initialization methods is based on Clifford circuits and Clifford parameter search. Clifford operations are a restricted subset of quantum gates that can be simulated efficiently on classical hardware while still preserving important structural properties of quantum states \cite{gottesman1998heisenberg}. In the context of QAOA, Clifford angles refer to special parameter settings for $\gamma$ and $\beta$ that cause parts of the quantum circuit to behave similarly to Clifford operations. These angles are useful because they provide stable and structured regions of the optimization landscape that can be explored efficiently before transitioning to fully variational optimization. Clifford-based search methods therefore attempt to identify promising initialization points by first exploring parameter settings that are easier to analyze and less sensitive to noise.

Several other initialization methods have been explored in the literature. Warm-start methods attempt to incorporate classical optimization solutions into the quantum initialization process~\cite{egger2021warm}, while layer-wise training methods progressively optimize deeper QAOA circuits using parameters learned from shallower circuits~\cite{zhou2020quantum}. Transfer-based initialization methods reuse parameters across related optimization instances to accelerate convergence. More recently, SPIQ \cite{bharadwaj2026spiq} has been proposed as a structured initialization approach that attempts to preserve information from the problem Hamiltonian when generating initial QAOA parameters. This makes SPIQ particularly attractive for combinatorial optimization problems such as join order optimization, where the search landscape is highly complex and sensitive to initialization quality.

While considerable research has been devoted independently to quantum formulations of join order optimization and to initialization strategies for variational quantum algorithms, little work has investigated the interaction between these two areas. To the best of our knowledge, SPIQ initialization has not previously been applied to quantum join order optimization, creating an opportunity to study whether structured initialization can improve convergence and solution quality for database optimization workloads.

\section{Method}

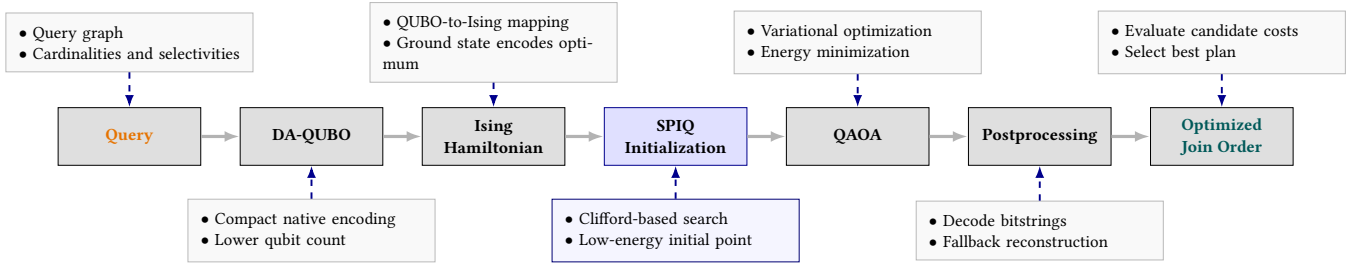
\begin{figure*}[t]
\centering
\resizebox{\textwidth}{!}{%
\begin{tikzpicture}[
    font=\small,
    box/.style={
        draw=black,
        fill=gray!25,
        minimum width=2.35cm,
        minimum height=0.95cm,
        align=center
    },
    note/.style={
        draw=gray!55,
        fill=gray!5,
        align=left,
        text width=3.65cm,
        inner sep=6pt,
        font=\small
    },
    arrow/.style={
        -{Latex[length=2.5mm,width=1.8mm]},
        line width=1.5pt,
        draw=gray!60
    },
    noteArrow/.style={
        -{Latex[length=2.0mm,width=1.5mm]},
        line width=0.9pt,
        draw=blue!55!black,
        dashed
    }
]

\node[box, text=orange!90!black] (query) at (0,0) {\textbf{Query}};
\node[box] (qubo) at (3.0,0) {\textbf{DA-QUBO}};
\node[box] (ising) at (6.0,0) {\textbf{Ising}\\\textbf{Hamiltonian}};
\node[box, fill=blue!12, draw=blue!55!black] (spiq) at (9.0,0) {\textbf{SPIQ}\\\textbf{Initialization}};
\node[box] (qaoa) at (12.0,0) {\textbf{QAOA}};
\node[box] (postprocessing) at (15.0,0) {\textbf{Postprocessing}};
\node[box, text=teal!70!black] (join) at (18.0,0) {\textbf{Optimized}\\\textbf{Join Order}};

\draw[arrow] (query) -- (qubo);
\draw[arrow] (qubo) -- (ising);
\draw[arrow] (ising) -- (spiq);
\draw[arrow] (spiq) -- (qaoa);
\draw[arrow] (qaoa) -- (postprocessing);
\draw[arrow] (postprocessing) -- (join);

\node[note] (n1) at (0,1.55) {
$\bullet$ Query graph\\
$\bullet$ Cardinalities and selectivities
};

\node[note] (n2) at (3.0,-1.55) {
$\bullet$ Compact native encoding\\
$\bullet$ Lower qubit count
};

\node[note] (n3) at (6.0,1.55) {
$\bullet$ QUBO-to-Ising mapping\\
$\bullet$ Ground state encodes optimum
};

\node[note, draw=blue!45!black, fill=blue!4] (n4) at (9.0,-1.55) {
$\bullet$ Clifford-based search\\
$\bullet$ Low-energy initial point
};

\node[note] (n5) at (12.0,1.55) {
$\bullet$ Variational optimization\\
$\bullet$ Energy minimization
};

\node[note] (n6) at (15.0,-1.55) {
$\bullet$ Decode bitstrings\\
$\bullet$ Fallback reconstruction
};

\node[note] (n7) at (18.0,1.55) {
$\bullet$ Evaluate candidate costs\\
$\bullet$ Select best plan
};

\draw[noteArrow] (n1.south) -- (query.north);
\draw[noteArrow] (n2.north) -- (qubo.south);
\draw[noteArrow] (n3.south) -- (ising.north);
\draw[noteArrow] (n4.north) -- (spiq.south);
\draw[noteArrow] (n5.south) -- (qaoa.north);
\draw[noteArrow] (n6.north) -- (postprocessing.south);
\draw[noteArrow] (n7.south) -- (join.north);

\end{tikzpicture}
}
\caption{Workflow of SPIQ-initialized QAOA for join order optimization.}
\label{fig:method_flow}
\end{figure*}

Our proposed approach begins by representing the join order optimization problem as a query graph containing relations, predicates, cardinalities, and predicate selectivities. We formulate the query graph as a QUBO using a compact state-of-the-art encoding proposed by Schonberger \cite{schonberger2023quantum} for Quantum Annealing approaches and adapt it to be used for our QAOA-based setup. The resulting QUBO is then transformed into an Ising Hamiltonian suitable for QAOA optimization. Before running QAOA, we apply SPIQ initialization to identify promising starting parameter values for the quantum circuit. SPIQ first performs a classical search over a simplified subset of parameter configurations using efficiently simulatable Clifford-based circuits, and then ranks candidate initialization points according to their energy values. The best-performing initialization point obtained from this search is subsequently used as the starting configuration for the full QAOA optimization process.  Figure~\ref{fig:method_flow} summarizes the overall pipeline, from query-graph construction and QUBO encoding to SPIQ initialization, QAOA optimization, sampling, and final join-order decoding. 



\subsection{QUBO Formulation}

As mentioned in Section 2, QAOA requires the
optimization problem to be expressed as a quadratic unconstrained
binary optimization (QUBO) problem, whose ground state corresponds
to the optimal solution. Join ordering is not naturally in this
form, so an encoding step is required. This encoding directly
determines the number of binary variables, and therefore the
number of qubits needed to represent the problem. The qubit count is the primary factor that bounds the size of solvable instances, since it scales as $2^n$ in the number of qubits. The choice of encoding
is therefore a central design decision.




We use the state-of-the-art native QUBO encoding proposed by Schonberger, Trummer, and Mauerer \cite{schonberger2023quantum}. Three design choices make this encoding better suited to join order optimization. The first is how the join tree is represented. It uses a single binary variable per relation per join, and enforces validity based on how many relations are active at each join and preventing a relation from leaving the tree once it has been joined.
The second difference lies in how cost is approximated. It uses a quadratic penalty on the logarithmic intermediate cardinality, which captures cost growth continuously. The third follows from the first two. With no inequality
constraints anywhere in the formulation, the mandatory variable count is only $(R + P)(J - 1)$ for $R$ relations, $P$ predicates, and $J$ joins.

To adapt this encoding to QAOA, we adopt the native encoding \cite{schonberger2023quantum}, which relies solely on the continuous quadratic cost approximation, so that no auxiliary variables are introduced beyond the (R+P)(J-1) qubits. The model is constructed with docplex  and translated, via qiskit-optimization's docplex translator, into a Qiskit QuadraticProgram.  the resulting cost term is the sum over joins of the squared logarithmic intermediate cardinality. We also attempted  implementing the MILP-based formulation \cite{schonberger2023ready}, but found it less suitable than the native encoding for QAOA, because its MILP-style inequality constraints require binary slack variables whose number grows as the JOO problem scales, increasing the qubit count well beyond the native encoding's (R+P)(J-1).

The QAOA optimization uses Qiskit's COBYLA optimizer. After performing a comparative study between the classical optimizers COBYLA, SPSA and AQGD, we found that COBYLA performed best with JOO while also achieving the fastest rate of convergence.




\subsection{Initialization with SPIQ}

A major challenge in QAOA is that the algorithm is highly sensitive to its starting parameters~\cite{cerezo2021variational}. In QAOA, these parameters control how the quantum circuit explores possible solutions during optimization. If the starting point is poor, the optimizer may become stuck in low-quality solutions or fail to improve the result meaningfully~\cite{mcclean2018barren}. In our join ordering experiments, random initialization often caused the optimizer to converge to suboptimal join plans or produce unstable optimization behavior. As a result, choosing good initial parameters is an important step in improving QAOA performance.

To address this issue, we integrate the Scalable Parameter Initialization for QAOA (SPIQ) framework~\cite{bharadwaj2026spiq} into our optimization pipeline. Instead of starting from random parameter values, SPIQ attempts to identify promising starting points before the full QAOA optimization begins. The main idea is to first explore a simplified version of the quantum circuit that can still capture useful structural information about the optimization problem while remaining computationally inexpensive to evaluate.

SPIQ performs this search using a restricted family of quantum operations known as Clifford operations \cite{gottesman1998heisenberg}. These operations are important because they can be simulated efficiently on classical hardware, unlike general quantum circuits which are typically very expensive to simulate. SPIQ searches over a small set of special parameter values, often referred to as Clifford angles, corresponding to rotations of $0$, $\frac{\pi}{2}$, $\pi$, and $\frac{3\pi}{2}$. Although these parameter settings represent only a simplified subset of all possible QAOA parameters, they still provide useful information about the structure of the optimization landscape.

To explore this parameter space efficiently, SPIQ uses a genetic algorithm~\cite{alam2020ga,gad2021pygad}. The search evaluates many candidate parameter settings and ranks them according to the energy of the resulting solution. Lower-energy states correspond to potentially better join orders. Rather than selecting parameters randomly, SPIQ therefore produces initialization points that are already biased toward promising regions of the optimization landscape.

\begin{figure}[h]
  \centering
  \includegraphics[width=\linewidth]{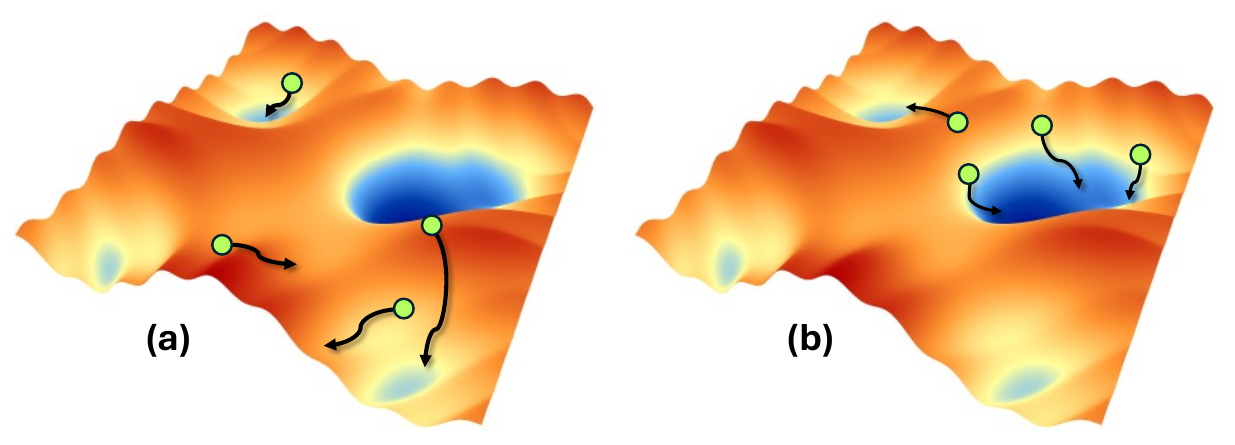}
  \caption{a) depicts points being trapped in sub-optimal or flat regions of the quantum landscape while b) illustrates landscape-aware points more proximal to superior solutions.}
  \label{fig:spiq}
\end{figure}

Finally, SPIQ selects a diverse set of high-quality initialization points for QAOA optimization. Instead of using only the single best candidate, the method attempts to choose candidates from different regions of the parameter space so that the optimizer does not repeatedly converge to the same local minimum. This improves robustness and increases the likelihood of discovering better join orders during optimization, as illustrated in Figure~\ref{fig:spiq}

The SPIQ initialization integrates into our join ordering workflow as a classical preprocessing stage between QUBO formulation and QAOA execution. The classical SPIQ phase adds no quantum resource overhead and scales polynomially with problem size. Overall, SPIQ provides a structured and problem-aware alternative to random initialization, making QAOA optimization more stable and effective for join order optimization problems.


 \subsection{Post-Processing}

 The post-processing stage first converts the output into a relation--join assignment matrix and computes a score vector that reflects how strongly each relation is associated with joins that occur earlier in the order. This ranked ordering is then converted into a candidate join plan whose actual execution cost is evaluated classically using standard join cardinality estimation.

We also employ the same fallback strategy introduced by Schonberger \cite{schonberger2023quantum} to handle cases where the output may lead to poor or invalid join structures, especially plans containing expensive cross products. Instead of relying entirely on the raw join order output, the fallback mechanism traverses the query graph while respecting the computed ranking order and deliberately avoids disconnected joins whenever possible. In practice, this means that the algorithm tries to construct a connected join sequence following the ranking produced while ensuring that predicates are used before introducing cross products. Finally, both the original candidate plan and the fallback plan are evaluated classically and the lower-cost plan is selected as the most optimal join order.

\section{Experimental Setup}
All experiments are executed using Qiskit 0.34.2\cite{qiskit2024} with the statevector simulator, which provides noiseless, exact quantum state evolution without shot-based sampling noise. The QUBO formulation stage uses Qiskit Optimization 0.3.2 and Gurobi 10.0.1\cite{gurobi} for constructing and validating the optimization model. The SPIQ initialization phase runs in a separate environment using Qiskit Aer 0.13.3 and Qiskit Algorithms 0.3.1 for stabilizer simulation and circuit construction. Since current gate-based quantum hardware imposes strict limits on qubit count, connectivity, and coherence time, simulation allows us to isolate the effects of encoding choice, optimizer selection, and initialization strategy from hardware-induced errors. All classical computation, including QUBO construction, SPIQ search, and QAOA parameter optimization, is performed on a workstation with an Intel Core i9-10940X CPU (14 cores, 28 threads, 3.30\,GHz) and 128\,GB of RAM.

\section{Preliminary Evaluation}

\subsection{Problem Instances}
We are limited to analyzing 3- and 4-table join orders due to resource constraints. To evaluate the impact of SPIQ initialization, we select the following join problems to cover a wide range of predicate selectivities, cardinalities, and cross-product presence:

\begin{enumerate}
    \item \textbf{Problem 1}: 3 relation join and 2 predicates with varied predicate selectivities. Cardinalities of the relations are 10, 15, and 20 respectively.There are 2 predicates, one between relations 1 and 2 with selectivity 0.1 and between relations 2 and 3 also with selectivity 0.1. Encoding this problem requires 6 qubits.
    \item \textbf{Problem 2}:  4 relation join and 2 predicates with low predicate connectivity and possible cross products. Cardinalities of the relations are 10, 15, and 20, 30 respectively. There are 2 predicates - between relations 1 and 2 and between relations 3 and 4, each with selectivity 0.1. Encoding this problem requires 12 qubits.
    \item \textbf{Problem 3}: 4 relation join and 6 predicates with high predicate connectivity and no cross products. Cardinalities of the relations are 10, 15, 20, and 30, respectively. There are predicates between every relation with selectivities ranging from 0.1 to 0.6. Encoding this problem requires 20 qubits.

\end{enumerate}

\subsection{Evaluation Metrics}

We evaluate performance along three axes:

\begin{itemize}
    \item \textbf{Cost ratio}: the fraction of sampled bitstrings that decode to join orders at each of the leading cost levels. For every configuration we report the sampling percentage of the top-3 most-sampled cost levels, ranked by cost, which reveals how probability mass is spread across join orders of differing cost.
    \item \textbf{Optimal ratio}: the fraction of sampled bitstrings that decode to the join order having least cost, compared between the first and final QAOA iterations. 
    \item \textbf{Energy convergence}: the trajectory of the expectation value over QAOA iterations, measuring how quickly and reliably optimizer reduces the energy and whether it stabilizes near the ground-state energy.
\end{itemize}

For each problem, we compare randomly initialized QAOA (random parameter initialization) against SPIQ-initialized QAOA under identical optimizer and iteration settings.

\subsection{Observations}

\begin{table}[htbp]
\centering
\small
\setlength{\tabcolsep}{3pt}
\renewcommand{\arraystretch}{0.95}
\caption{Top-three final-state cost distributions of evaluation problems}
\label{tab:top3_problem_cost_distributions}
\resizebox{\columnwidth}{!}{%
\begin{tabular}{l|rr|rr|rr|rr}
\toprule
\multirow{3}{*}{Problem}
& \multicolumn{4}{c|}{Randomly Initialized}
& \multicolumn{4}{c}{SPIQ Initialized} \\
\cmidrule(lr){2-5}
\cmidrule(lr){6-9}
& \multicolumn{2}{c|}{fallback=false}
& \multicolumn{2}{c|}{fallback=true}
& \multicolumn{2}{c|}{fallback=false}
& \multicolumn{2}{c}{fallback=true} \\
\cmidrule(lr){2-3}
\cmidrule(lr){4-5}
\cmidrule(lr){6-7}
\cmidrule(lr){8-9}
& Cost & pct
& Cost & pct
& Cost & pct
& Cost & pct \\
\midrule

\multirow{3}{*}{P1}
& \textbf{15}  & \textbf{12.1\%}
& \textbf{15}  & \textbf{46.3\%}
& \textbf{15}  & \textbf{49.3\%}
& \textbf{15}  & \textbf{49.4\%} \\
& 30  & 56.0\%
& 30  & 53.7\%
& 30  & 50.6\%
& 30  & 50.6\% \\
& 200 & 31.8\%
& --  & --
& 200 & 0.127\%
& --  & -- \\
\midrule

\multirow{3}{*}{P2}
& \textbf{900}  & \textbf{18.2}\%
& \textbf{465}  & \textbf{23.2}\%
& \textbf{465}  & \textbf{50.0}\%
& \textbf{465}  & \textbf{50.0}\% \\
& 960  & 20.4\%
& 660  & 24.4\%
& 600  & 24.5\%
& 960  & 50.0\% \\
& 1350 & 10.6\%
& 960  & 42.5\%
& 960  & 25.4\%
& --   & -- \\
\midrule

\multirow{3}{*}{P3}
& \textbf{440} & \textbf{7.81}\%
& \textbf{39}  & \textbf{14.1}\%
& \textbf{39}  & \textbf{50.0}\%
& \textbf{39}  & \textbf{74.5}\% \\
& 510 & 11.7\%
& 104 & 14.9\%
& 84  & 25.5\%
& 84  & 0.0391\% \\
& 600 & 8.41\%
& 375 & 25.9\%
& 240 & 24.4\%
& 104 & 25.5\% \\
\bottomrule
\end{tabular}
}
\end{table}

\begin{figure}[htbp]
    \centering
    \includegraphics[width=1\linewidth]{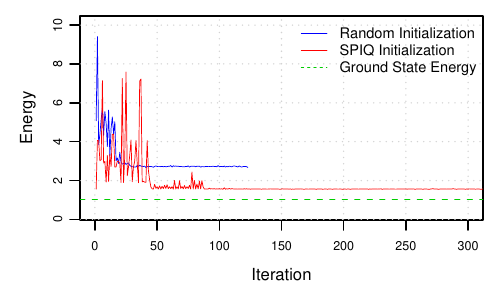}
    \caption{ Energy convergence for Problem 1}
    \label{fig:3table2pre}
\end{figure}

\begin{figure}[htbp]

    \centering

    \includegraphics[width=1\linewidth]{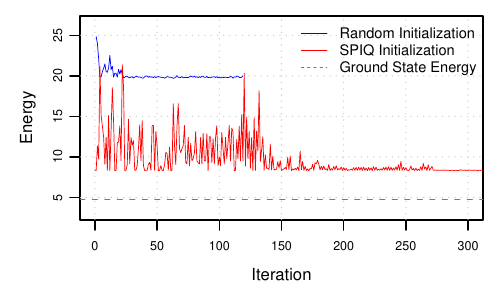}

    \caption{Energy convergence for Problem 2}

    \label{fig:4table2pred}

\end{figure}

\begin{figure}[htbp]
    \centering
    \includegraphics[width=1\linewidth]{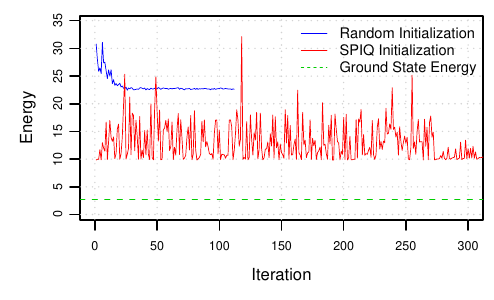}
    \caption{ Energy convergence for Problem 3}
    \label{fig:4table6pred}
\end{figure}

Table~\ref{tab:top3_problem_cost_distributions} reports, for each of the three problems, the three most-sampled final-state cost levels under four settings: random initialization without fallback, random initialization with fallback, SPIQ initialization without fallback, and SPIQ initialization with fallback. Because only the top-3 sampled costs are listed, the table characterizes where the bulk of the sampled probability mass concentrates. Across these reported costs, SPIQ initialization generally shifts probability mass toward lower cost outcomes relative to random initialization , increasing the share of the optimal join order among the most-sampled levels.


With fallback disabled, the shift is clearest in Problems~2 and~3. Problem~1 improves more modestly: SPIQ achieves a 4$\times$ improvement optimal cost frequency and nearly eliminates the high cost level~$200$ (from $31.8\%$ to $0.127\%$), but cost~$30$ remains comparably sampled ($50.6\%$), so the optimal order becomes prominent but not clearly dominating the reported costs. 

With fallback enabled, SPIQ again increases the share of the optimal cost among the reported top 3 levels, with the magnitude differing by problem. The effect is strongest in Problem~3, where
SPIQ achieves a 5$\times$ improvement in optimal cost frequency. In In Problem~2, SPIQ achieves a 2.2$\times$ improvement in optimal cost frequency($23.2\%$ to $50.0\%$.). Problem~1 shows the mildest change, consistent with its small search space. Taken together, these results suggest that SPIQ initialization tends to concentrate the most-sampled costs on lower cost join orders.

As shown in Figures~\ref{fig:3table2pre} and~\ref{fig:4table2pred}, our experimental results show that the energies obtained by QAOA do not yet reach the theoretical ground state, indicating that there is still significant room for improvement in optimization quality, parameter tuning, and hardware scalability. Although SPIQ-initialized QAOA converges to energy values closer to the theoretical ground state than randomly initialized QAOA, the decrease from the initial SPIQ energy to the converged energy is relatively limited. This suggests that SPIQ provides a better starting point, but the optimization process itself does not substantially improve the solution.


\section{Insights and Future Work}
Our results indicate that SPIQ initialization consistently produces solutions with energies closer to the theoretical ground state than randomly initialized QAOA, suggesting that structured initialization improves optimization stability and guides the algorithm toward higher-quality join plans. However, the energy reduction achieved during the variational optimization itself is limited, as the join ordering landscape remains difficult to optimize in practice.

This points to two complementary directions for future work. On the optimization side, the primary goal is to improve QAOA's ability to descend below the SPIQ starting point; we are hopeful that this trend will emerge as we move to more complex JOO problems where the variational phase has more room to contribute. On the initialization side, since the starting point is where most of the benefit currently arises, it is worth investigating whether more advanced initialization strategies can further sharpen the starting distribution and ease the subsequent optimization.


\begin{acks}
This work was supported (in part) by the Google ML and Systems Junior Faculty Award, the Google JAX AI Stack Research Award, and the NVIDIA Academic Grant Program Award.
This material is based upon work supported by the U.S. Department of Energy, Office of Science, Office of Advanced Scientific Computing Research, Accelerated Research in Quantum Computing under Award Number DE-SC0025633. 
This research was, in part, funded by the U.S. Government.  The views and conclusions contained in this document are those of the authors and should not be interpreted as representing the official policies, either expressed or implied, of the U.S. Government.
\end{acks}

\clearpage

\bibliographystyle{ACM-Reference-Format}
\bibliography{main}

\end{document}